\documentclass[conference]{IEEEtran}
\usepackage{cite}
\usepackage{amsmath,amssymb}
\usepackage{graphicx}
\usepackage{booktabs}
\usepackage{xcolor}
\usepackage{url}

\begin{document}

\title{Energy and Performance Characterization of UxNBs: Toward AI-Driven Non-Terrestrial Open RAN Control}

\author{\IEEEauthorblockN{Ali Srour, Farzad Veisi, Sami Taktak, Vania Conan}
\IEEEauthorblockA{\textit{CEDRIC Lab, CNAM}, Paris, France \\
\{ali.srour, farzad.veisi-goshtab, sami.taktak, vania.conan\}@cnam.fr}}

\maketitle
\begin{abstract}
UAV-mounted base stations (UxNBs) can provide on-demand 5G coverage, but
their limited onboard energy makes power consumption the main operational
constraint. AI-driven Open RAN controllers are a promising solution, yet
they are trained almost exclusively on analytical models and simulations,
since no public dataset captures the real energy behavior of a flying 5G
base station (BS). As a first step toward closing this gap, we present a UAV
testbed hosting a 5G picocell that supports a 20\,MHz carrier and up to
100\,Mbps throughput. The testbed measures BS system metrics, including power consumption, CPU load, and temperature, alongside UAV flight telemetry and detailed per-layer vRAN processing metrics. To our knowledge, this is the first demonstration that jointly exposes the
energy and RAN behavior of a flying 5G BS, providing the real
measurements needed to train and validate energy-aware AI controllers.
\end{abstract}
\begin{IEEEkeywords}
UxNB, UAV base station, 5G, Open RAN, energy efficiency, dataset
\end{IEEEkeywords}

\section{Introduction}

Unmanned aerial vehicles carrying a radio access node, referred to as UxNBs
\cite{3GPP_2019}, can deliver cellular coverage where terrestrial
infrastructure is missing, overloaded, or damaged. Several
prototypes have shown that a full cellular stack can operate on board a UAV,
from early LTE systems \cite{Conext_2018}
to recent 5G aerial BSs built on open-source software and commodity
hardware \cite{WCNC_2024}. Energy is the main obstacle to operating UxNBs, since one battery powers both
propulsion and the BS payload. Although propulsion usually dominates, the UxNB payload is far from
negligible: an aerial 5G DU on commodity hardware draws about 28\,W under
traffic \cite{arXiv_2026a}, and its power varies with load. Unlike
propulsion, this share is controllable by the RAN, making it a natural
target for intelligent control.

The Open RAN architecture enables such control through AI-driven xApps and
rApps hosted on its RAN Intelligent Controllers.
Learning-based controllers already reduce the energy of terrestrial and
virtualized BSs \cite{INFOCOM_2021},
and similar approaches have been proposed to optimize the placement, trajectory,
and resources of UAV BSs \cite{TCC_2023}.
However, the terrestrial solutions are designed for ground cells, while the UAV
solutions are evaluated in simulation on analytical power models. One limiting factor is the scarcity of real measurement data. Public UAV flight energy traces exist,
but without a BS on board, public base
station power measurements are collected on the ground
\cite{ICC_2021}, and the only public UAV BS power
dataset we found is computed from analytical models \cite{Zenodo_2024}.
Meanwhile, UxNB prototypes report network performance but release no detailed energy
traces across different parts of UxNB. This gap matters because learned network controllers that perform well
in emulation often fail to deliver the same gains on real systems
\cite{NSDI_2020}.

As a first step toward closing this gap, this demo presents a UAV-based 5G
testbed for the energy and performance characterization of UxNBs. The testbed
hosts a 5G picocell with a 20\,MHz carrier and up to 100\,Mbps throughput,
built from commodity components, and jointly records BS system
metrics, UAV flight telemetry, and RAN performance metrics. By showing these
metrics side by side, the demo provides real data on which energy-aware AI
controllers for non-terrestrial Open RAN can be designed and validated.

\section{Testbed and Methodology}

\begin{figure}[htbp]
    \centering
    \includegraphics[width=0.95\linewidth]{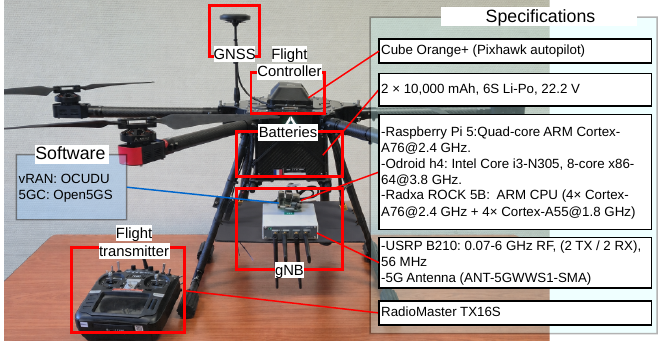}    
    \vspace{-2mm}

  \caption{UxNB testbed hardware.}
    \label{fig:testbed}
\end{figure}
\begin{figure}[t]
    \centering
    \includegraphics[width=0.82\linewidth, trim=0 6 0 0, clip]{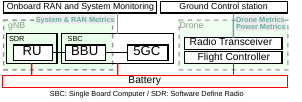}
    \vspace{-2mm}
    \caption{Testbed architecture and monitored metric groups.}
    \label{fig:arch}
    \vspace{-2mm}
\end{figure}
   
\textbf{UAV platform.} The UxNB is built on an Aurelia X6 hexacopter running
the ArduPilot autopilot on a Cube Orange+ flight controller with GNSS, and
powered by two 10{,}000\,mAh 6S Li-Po batteries (22.2\,V), as shown in
Fig.~\ref{fig:testbed}. The platform carries up to 5\,kg of payload, with a
flight time of about 30\,min without payload and 13\,min at full payload. A
ground control station running Mission Planner on a separate machine controls
the flight and receives the flight telemetry. 

\textbf{gNB payload.} The gNB follows the split of Fig.~\ref{fig:arch}: a USRP
B210 SDR acts as the radio unit, while a single-board computer (SBC) runs the
baseband processing with the OCUDU vRAN stack and the Open5GS 5G core. We test three SBCs: two ARM boards, a Raspberry Pi~5 and a Radxa ROCK~5B, and one x86 board, an Odroid~H4, to compare how the compute platform type affects energy and performance. The cell operates on a 20 MHz carrier in band n77 and delivers up to 100 Mbps to two commercial UEs, depending on the processing resources allocated to the gNB. The gNB hardware weighs less than 1 kg. Since both the gNB and the 5G core run on board, the UxNB operates as a network-in-a-box that needs no backhaul to an external core
\cite{MobiCom_2018}.

\textbf{Monitoring system.} A custom lightweight monitoring system collects the
gNB metrics with a minimal footprint, since on a constrained processing unit
any monitoring overhead competes with the vRAN for CPU time and can cause
processing deadline violations and data drops, especially in the physical
layer \cite{GLOBECOM_2026}. Power consumption is measured by the drone's power
module and reported by the flight controller to the ground station, together
with the flight metrics. All metrics share a common time reference; the main
ones are listed in Table~\ref{tab:metrics}, and the full list of
metrics and the collection scripts are available at \url{https://gitlab.roc.cnam.fr/srour/uxnb-testbed}. \textbf{Monitored metrics.} Beyond CPU utilization and memory, the system
records processing overhead indicators, such as context switches, thread
counts, and instruction, cycle, and cache-miss counts, which reveal how
efficiently a constrained board executes the vRAN. The system also monitors the fronthaul
interface between the processing unit and the SDR, which carries a continuous
I/Q sample stream, and the backhaul link that carries user traffic,
since on such small platforms these interfaces can limit throughput before the
CPU does. Within the vRAN, processing metrics are
collected per layer, from per-channel PHY processing times and task queue
depths up to the data volume handled by SDAP, PDCP, and RLC, so that the
impact of traffic load, channel conditions, and configuration can be traced to
the layer where it arises.

\textbf{Methodology.} End-to-end downlink and uplink tests sweep the offered
load up to the maximum achievable throughput, which is bounded either by the
20\,MHz bandwidth or by the processing capacity of the SBC, since each board
reaches a different data rate depending on its CPU and interface capability.
Tests consider two scenarios in which the drone is stationary and serves two
UEs, first static and then moving, under different traffic profiles. As a first result,
at 50\,Mbps the BBU power reaches 18\,W on the Odroid,
8\,W on the Pi~5, and 7.6\,W on the Radxa, i.e., 12.5\%, 31.2\%, and
27\% above idle.
\begin{table}[t]
\centering
\caption{Monitored Metrics}
\label{tab:metrics}
\scriptsize
\setlength{\tabcolsep}{2pt}
\renewcommand{\arraystretch}{1.15}
\resizebox{\columnwidth}{!}{%
\begin{tabular}{@{}ll@{}}
\toprule
\textbf{Category} & \textbf{Metrics} \\
\midrule
\multicolumn{2}{@{}l@{}}{\textbf{Power metrics}} \\
Battery & Voltage, current, power, consumed capacity (BBU, RU, Propulsion) \\
\midrule
\multicolumn{2}{@{}l@{}}{\textbf{Drone metrics}} \\
Flight & Position, altitude, speed, attitude, motor outputs, GPS status \\
\midrule
\multicolumn{2}{@{}l@{}}{\textbf{System metrics}} \\
CPU/System & Frequency, utilization, RSS, threads, context switches, cycles, cache misses \\
Temperature & CPU/SoC temperature, thermal throttling state \\
Interfaces & Fronthaul and backhaul link data rate \\
\midrule
\multicolumn{2}{@{}l@{}}{\textbf{RAN metrics}} \\
Processing time & PHY (PDSCH, PUSCH, PDCCH, PUCCH) time per slot \\
Buffers & RLC, MAC DL/UL, PUSCH, PHY, FAPI, HARQ \\
HARQ & ACK/NACK, retransmissions, discards, timeouts \\
Radio/RF & Late TX, underflows, overflows \\
Link quality & CQI, RI, MCS, BLER, SNR, PRBs, queue latency \\
E2E Throughput & DL/UL rate, throughput, jitter, packet loss \\
\bottomrule
\end{tabular}%
}
\end{table}
\section{Demonstration}

The demonstration presents recorded videos of the two
stationary-drone scenarios described in Section~II, with the recorded metrics
replayed on a dashboard synchronized with the videos. The dashboard shows the power consumption of the
three parts of the system: the vRAN processing, the radio unit, and the drone.
Alongside the power, it displays the system and RAN metrics that drive it,
namely CPU utilization, allocated PRBs, MCS, and BLER, together with the
end-to-end throughput, so that attendees can see how traffic and UE mobility
affect both energy and performance. Finally, the dashboard compares the
three processing boards, highlighting the differences in power and achievable
throughput between the ARM and x86 platforms.

\section{Conclusion and Future Work}

This demo presented a UAV-based 5G testbed that jointly records the energy,
compute, and RAN behavior of a UxNB on three processing platforms. As future
work, we will extend the measurement campaign and release the collected data
as an open dataset, which will serve to train and validate energy-aware AI
controllers, such as xApps and rApps, for non-terrestrial Open RAN.
\section*{Acknowledgements}

This work was funded by the France 2030 IE6 (Internet of the Edges for 6G; contract nb DOS0223931) and IPCEI ME/CT (contract nb DOS0239248/00) projects.

\bibliographystyle{IEEEtran}
\bibliography{references}

@techreport{3GPP_2019,
  author    = {{3GPP}},
  title     = {Enhancement for Unmanned Aerial Vehicles ({UAVs}); Stage 1},
  institution = {3GPP},
  number    = {TR 22.829},
  year      = {2019}
}

@inproceedings{Conext_2018,
  author    = {Chakraborty, A. and others},
  title     = {{SkyRAN}: A Self-Organizing {LTE} {RAN} in the Sky},
  booktitle = {ACM CoNEXT},
  year      = {2018}
}

@inproceedings{MobiCom_2018,
  author    = {Moradi, M. and others},
  title     = {{SkyCore}: Moving Core to the Edge for Untethered and Reliable {UAV}-based {LTE} Networks},
  booktitle = {ACM MobiCom},
  year      = {2018}
}

@inproceedings{WCNC_2024,
  author    = {Mishra, D. and others},
  title     = {{SKY5G}: Prototyping {5G} Aerial Base Station ({UAV-BS}) for On-Demand Connectivity from Sky},
  booktitle = {IEEE WCNC},
  year      = {2024}
}

@misc{arXiv_2026a,
  author    = {Duboc, M. and others},
  title     = {Toward Lightweight Aerial {5G} {gNBs}: Reproducible {OAI} Testbed on Commodity {ARM} Platforms},
  howpublished = {arXiv},
  year      = {2026}
}

@article{TCC_2023,
  author    = {Pham, C. and others},
  title     = {When {RAN} Intelligent Controller in {O-RAN} Meets Multi-{UAV} Enable Wireless Network},
  journal   = {IEEE TCC},
  year      = {2023}
}

@inproceedings{INFOCOM_2021,
  author    = {Ayala-Romero, J. A. and others},
  title     = {Bayesian Online Learning for Energy-Aware Resource Orchestration in Virtualized {RANs}},
  booktitle = {IEEE INFOCOM},
  year      = {2021}
}

@inproceedings{ICC_2021,
  author    = {Ayala-Romero, J. A. and others},
  title     = {Experimental Evaluation of Power Consumption in Virtualized Base Stations},
  booktitle = {IEEE ICC},
  year      = {2021}
}

@misc{Zenodo_2024,
  author       = {{6G-DATADRIVEN}},
  title        = {Power Balance Characteristics for Multirotor- and Fixed-Wing-Type {UAV-BSs} Equipped with {RES} and {RISs}},
  howpublished = {Zenodo},
  year         = {2024}
}

@inproceedings{NSDI_2020,
  author    = {Yan, F. Y. and others},
  title     = {Learning in Situ: A Randomized Experiment in Video Streaming},
  booktitle = {USENIX NSDI},
  year      = {2020}
}

@inproceedings{GLOBECOM_2026,
  author    = {Srour, A. and others},
  title     = {{ALARM}: Adaptive Layer-Aware Resource Management for Power-Efficient {vRANs}},
  booktitle = {IEEE GLOBECOM},
  year      = {2026}
}

\end{document}